\documentclass[aps,prb,twocolumn,superscriptaddress,showpacs]{revtex4}

\usepackage{amsmath,amssymb,latexsym}
\usepackage{graphicx}
\graphicspath{{../fig/}}
\usepackage[dvipdfm,hyperfootnotes=false]{hyperref}
\usepackage{txfonts}
\usepackage{mathrsfs}
\usepackage{color}
\DeclareSymbolFont{symbols}{OMS}{cmsy}{m}{n}
\DeclareSymbolFont{largesymbols}{OMX}{cmex}{m}{n}
\newcommand{\bm}[1]{\boldsymbol #1}

\begin{document}

\title{
Theory of Anderson pseudospin resonance with Higgs mode in superconductors
%driven by ac electric fields
}

\author{Naoto Tsuji}
\affiliation{Department of Physics, University of Tokyo, Hongo, Tokyo 113-0033 , Japan}
\author{Hideo Aoki}
\affiliation{Department of Physics, University of Tokyo, Hongo, Tokyo 113-0033 , Japan}
%\email[]{}
%\homepage[]{}
%\thanks{}
%\altaffiliation{}

\begin{abstract}
A superconductor illuminated by an ac electric field with frequency $\Omega$ is
theoretically found to generate
a collective precession of Anderson's pseudospins, and hence a coherent amplitude oscillation of the order parameter,
with a doubled frequency $2\Omega$ through a nonlinear light-matter coupling.  
We provide a fundamental theory, based on the mean-field formalism,
to show that 
the induced pseudospin precession resonates with the Higgs amplitude mode of the superconductor at $2\Omega=2\Delta$ with $2\Delta$ being 
the superconducting gap. 
%We propose to employ Anderson's pseudospin representing superconductors to describe a resonance 
%phenomenon where 
%which is analogous to nuclear magnetic resonance for a nuclear spin in an oscillating magnetic field. 
%We provide a fundamental theory for Anderson pseudospin resonance,
%and show that the resonance, 
%Higgs?, 
%occurs at $2\Omega=2\Delta$ 
%in the subgap regime ($\Omega$ smaller than the superconducting gap 
%$2\Delta$). 
The resonant precession is accompanied by a divergent enhancement of the third-harmonic generation (THG).
By decomposing the THG susceptibility into the bare one and vertex correction,
we find that the enhancement of the THG cannot be explained by individual quasiparticle excitations (pair breaking),
so that the THG serves as a smoking gun for an identification of the collective Higgs mode.
We further explore the effect of electron-electron scattering on the pseudospin resonance
by applying the nonequilibrium dynamical mean-field theory to the attractive Hubbard model driven by ac electric fields.
The result indicates that the pseudospin resonance is robust against electron correlations,
although the resonance width is broadened due to electron scattering, which determines the lifetime of the Higgs mode.
\end{abstract}

%\collaboration{}
%\noaffiliation

\date{\today}


\maketitle

\section{Introduction}

Dynamical control of quantum many-body states of matter without destroying quantum coherence 
%by external perturbations such as laser irradiation 
is becoming a central challenge in condensed matter physics.
While recent developments in ultrafast laser experiments have enabled one to
study relaxation dynamics of quantum systems after pulse excitation, 
%@it is rather difficult to predict 
%the states realized at a time well after the pulse.  Alternatively, 
an alternative direction we can pursue is to look at far-from-equilibrium quantum states 
that are realized {\it during photoirradiation}. 
%The key difficulty is how to preserve quantum coherence with a continuous drive.
%For example, when particles are coupled to a time-periodic electric field,
%they can effectively dress a photon field, and turn into quasiparticles
%(a so-called Floquet state), whose energy spectrum is modified due to the coupling to the electric field.

From this viewpoint, superconductivity is an intriguing ground 
%among quantum many-body states 
to look for a novel optical control.
%With the gauge symmetry spontaneously broken, the phase mode of a superconductor is 
%obviously interesting. 
A superconducting state can be described in terms of pseudospins introduced by Anderson in 1958.\cite{Anderson1958}
Indeed, a collective precession of the pseudospins 
represents a Higgs amplitude mode \cite{Anderson1958,VolkovKogan1973,LittlewoodVarma1981a,Varma2002,PekkerVarma2015}, 
%VolovikZubkov2014}
i.e., a coherent amplitude oscillation of the superconducting order parameter with a frequency $2\Delta$ (the superconducting gap), which is a condensed matter analog of the Higgs boson in
elementary particle physics,\cite{EnglertBrout1964,Higgs1964,Guralnik1964} and the $\sigma$ meson in nuclear physics.\cite{NambuJonaLasinio1961}
This naturally emerges as a massive mode along the radial direction in the Mexican-hat potential profile
when a spontaneous symmetry breaking occurs in systems coupled to gauge fields \cite{Anderson1963,EnglertBrout1964,Higgs1964,Guralnik1964}.
%Namely, the amplitude of the superconducting order parameter 
%coherently oscillates with a frequency $2\Delta$ (the superconducting gap), 
The Higgs mode in superconductors has been experimentally observed by Raman \cite{SooryakumarKlein1980,Measson2014}
and THz pump-probe\cite{Matsunaga2013} spectroscopies.  
A natural question then is whether one can manipulate the dynamics of the pseudospins like one does for real spins by applying a magnetic field.  
Usually, however, 
it has been supposed to be difficult to photo-control the pseudospins,
%this is hindered, 
since the pseudospins do not directly couple to electromagnetic fields 
(in the linear-response regime).
%so that it has been supposed difficult to photo-control the pseudospins.

In this paper, we theoretically show that, 
if we go over to a {\it nonlinear} regime, 
%we pursue such a possibility for superconductivity,
an ac electric field with frequency $\Omega$ does indeed generate a collective precession of Anderson's pseudospins
%representing the superconducting order parameter
with frequency $2\Omega$ through the nonlinear light-matter coupling,
which results in a $2\Omega$ amplitude oscillation of the superconducting order parameter.
We further find that a {\it resonance} between the induced pseudospin precession and the Higgs mode emerges when $2\Omega=2\Delta$. 
This is remarkable, since 
%one might first think that the resonance should occur at 
%$\Omega=2\Delta$ when the photon energy coincides with the coherence peak-to-peak distance in the density of states of a superconductor.  
%In fact, 
this occurs {\it not} at $\Omega=2\Delta$ but at $\Omega$ well below $2\Delta$ (subgap regime), 
where quasiparticle excitations 
%that disturb superconducting coherence 
are suppressed.  
%A striking consequence of the collective excitation 
%is found to be a divergence of the amplitude of the induced pseudospin precession.
We may call the phenomenon ``Anderson pseudospin resonance" (APR).
APR may seem analogous to the nuclear magnetic resonance (NMR) or electron spin resonance (ESR), but APR is distinct in that the effect is essentially a {\it collective phenomenon as a resonance with the Higgs amplitude mode.} 
We show that APR should appear as a divergent enhancement of the third-order nonlinear optical response [third harmonic generation (THG)].
We further find that the enhancement of THG cannot be explained by quasiparticle excitations, which hence distinguishes the collective Higgs mode from individual pair breaking processes,
both of which lie at the same energy scale.
APR has been experimentally observed very recently by a THz laser experiment \cite{Matsunaga2014}.

\section{Phenomenological time-dependent Ginzburg-Landau theory}

To understand how the order parameter and the Higgs amplitude mode dynamically respond
to electromagnetic fields,
it is instructive to first overview the time-dependent Ginzburg-Landau (GL) theory.
This gives a simple macroscopic (and phenomenological) description of the superconducting order parameter as a low-energy effective field theory, although we have to mention that the time-dependent GL theory has a serious problem in describing the Higgs mode and its resonance
in superconductors as we shall stress toward the end of this section, which makes us opt for a microscopic theory in later sections.

Let us consider the GL ``Lagrangian density'' as a functional of the complex
order parameter $\Psi(\bm r,t)$ in a general form of
%which includes all the terms that are relevant at low energy and that are allowed by symmetry.
%several issues related to whether it can be justified from a microscopic theory or not.
%Strictly speaking, a natural form of the time-dependent GL theory that we present in this section 
%and is often employed to analyze experiments is
%not correct for clean and impure superconductors with a finite gap
%from the microscopic point of view. We give the complete microscopic results in later sections.
%Nevertheless, we feel that it is worthwhile to get
%a physical intuition for the Higgs resonance phenomenon from this rather simplified view.
%The GL free energy is a functional of the spatially varying order parameter $\Psi(\bm r)$,
%which takes a form of
\begin{align}
\mathcal L
&=
-\left[a|\Psi|^2+\frac{b}{2}|\Psi|^4+\frac{1}{2m^\ast}|(-i\nabla-e^\ast \bm A)\Psi|^2\right],
\nonumber
\\
&\quad
+c|(i\partial_t-e^\ast\phi)\Psi|^2+d\Psi^\dagger(i\partial_t-e^\ast\phi)\Psi
%+\gamma\Psi^\dagger \partial_t\Psi.
\label{GL Lagrangian}
\end{align}
where $a, b, c$ and $d$ are coefficients,
$\phi$ and $\bm A$ are the scalar and vector potentials, and
$e^\ast$ and $m^\ast$ are the effective electric charge and effective mass, respectively.
The Lagrangian density (\ref{GL Lagrangian}) is invariant under the gauge transformation $\Psi(\bm r,t)\to e^{ie^\ast\chi(\bm r,t)}\Psi(\bm r,t)$,
$\phi(\bm r,t)\to \phi(\bm r,t)-\partial_t\chi(\bm r,t)$, 
$\bm A(\bm r,t)\to \bm A(\bm r,t)+\nabla\chi(\bm r,t)$.
%It respects the global $U(1)$ phase-rotation symmetry ($\Psi\to e^{i\theta}\Psi$).
At temperatures $T<T_c$, $a=a_0(T-T_c)$ becomes negative, and 
the global $U(1)$ symmetry [$\Psi(\bm r,t)\to e^{ie^\ast\chi}\Psi(\bm r,t)$ with a constant $\chi$] is spontaneously broken.
The other coefficients are taken to be positive.
To describe the dynamics of the order parameter, we have included the kinetic terms,
one with a coefficient $c$ that represents the kinetic term of Klein-Gordon-type equations,
and another with $d$ that represents the kinetic term of Gross-Pitaevskii-type equations.
%the first and second time derivatives.
%$c$ and $d$ terms are kinetic terms of the Klein-Gordon and Gross-Pitaevskii 
%equations of motion, respectively.
%while $\gamma$ term represents damping or relaxation of the order parameter.
%$d$ term provides the Gross-Pitaevskii-type equation of motion.

Now, we expand $\mathcal L$ (\ref{GL Lagrangian}) around the ground state $\Psi_0=\sqrt{-a/b}$
(the phase is chosen as such without loss of generality).
There are two kinds of elementary excitations from the ground state: the variation along the radial direction
and another along the circumferential direction
on the complex plane of the order parameter. We write them as
$\Psi(\bm r,t)=[\Psi_0+H(\bm r,t)]e^{i\theta(\bm r,t)}$, where $H$ and $\theta$ denote the Higgs and 
Nambu-Goldstone (NG) fields, respectively.
The expansion gives us
\begin{align}
\mathcal L
&=
c(\partial_t H)^2+ce^\ast{}^2\left(\phi+\frac{1}{e^\ast}\partial_t\theta\right)^2(\Psi_0+H)^2
\nonumber
\\
&\quad
-de^\ast\left(\phi+\frac{1}{e^\ast}\partial_t\theta\right)(\Psi_0+H)^2
+2aH^2
-\frac{1}{2m^\ast}(\nabla H)^2
\nonumber
\\
&\quad
-\frac{e^\ast{}^2}{2m^\ast}
\left(\bm A-\frac{1}{e^\ast}\nabla\theta\right)^2(\Psi_0+H)^2
+\cdots
%-\mathcal L
%&\sim
%\begin{pmatrix}
%h & \phi
%\end{pmatrix}
%\begin{pmatrix}
%c\partial_t^2-2a-\frac{\nabla^2}{2m^\ast} & d\partial_t \\
%-d\partial_t & c\partial_t^2-\frac{\nabla^2}{2m^\ast}
%\end{pmatrix}
%\begin{pmatrix}
%h \\ \phi
%\end{pmatrix}
%\nonumber
%\\
%&\quad
%-\frac{e^\ast \Psi_0}{m^\ast}\bm A\nabla\phi
%+\frac{e^\ast}{m^\ast}\bm A(\phi\nabla h-h\nabla\phi)
%\nonumber
%\\
%&\quad
%+\frac{e^\ast{}^2\Psi_0}{m^\ast}\bm A^2 h
%+\cdots,
\label{GL expanded}
\end{align}
in which we have dropped total-derivative terms as well as
higher-order interactions. 
%\textcolor{red}{
%Note that the Higgs field is coupled to the NG field in the forms of
%$(\partial_t\theta) H$ and $(\bm A\cdot\nabla\theta) H$.
%}

The terms proportional to $\phi\partial_t\theta$ and $\bm A\cdot\nabla\theta$ in Eq.~(\ref{GL expanded}) indicate
that the NG phase mode turns into a longitudinal component of the gauge field.
As a result of the Anderson-Higgs mechanism,\cite{Anderson1963,EnglertBrout1964,Higgs1964,Guralnik1964} the NG mode is 
absorbed to the gauge field, and is pushed to very high energy scale of the plasma frequency $\omega_p$. 
We can thus regard $\theta$ in Eq.~(\ref{GL expanded})
to be an unphysical degree of freedom, which one can eliminate by taking the unitary gauge,
\begin{align}
\mathcal L
&=
c(\partial_t H)^2
+(ce^\ast{}^2\phi^2-de^\ast\phi)(\Psi_0^2+2\Psi_0 H)
+2aH^2
\nonumber
\\
&\quad
-\frac{1}{2m^\ast}(\nabla H)^2
-\frac{e^\ast{}^2\Psi_0^2}{2m^\ast}\bm A^2+\frac{e^\ast{}^2\Psi_0}{m^\ast}\bm A^2 H
+\cdots.
%-\mathcal L
%\sim
%h\left(c\partial_t^2-2a-\frac{\nabla^2}{2m^\ast}\right)h
%+\frac{e^\ast{}^2\Psi_0}{m^\ast}\bm A^2 h.
\label{GL higgs}
\end{align}
One can see that the terms $ce^\ast{}^2\phi^2$ and $\frac{e^\ast{}^2\Psi_0^2}{2m^\ast}\bm A^2$
represent the mass of the gauge field generated via the Anderson-Higgs mechanism.
%Hence the effective theory for the Higgs field becomes ``relativistic'' (with an emergent Lorentz symmetry),\cite{PekkerVarma2015}
%meaning that the first time derivative is absent even though we started from the non-relativistic GL
%Lagrangian (\ref{GL Lagrangian}).

In the case of electrically neutral superfluids ($e^\ast=0$), the Anderson-Higgs mechanism
does not occur, so that the Higgs field mixes with the NG field
via the term proportional to $d(\partial_t\theta)H$
in Eq.~(\ref{GL expanded}), and the Higgs mode is no longer considered to be an isolated excitation.
Furthermore, there are interactions between $H$ and $\theta$
via the terms proportional to $c(\partial_t\theta)^2 H$ and $(\nabla\theta)^2 H$
in Eq.~(\ref{GL expanded}), 
which causes the relaxation of the Higgs into lower energy NG bosons,
which makes the Higgs mode unstable.
At this point, it has been often emphasized that 
the particle-hole symmetry is important in
forcing $d\sim 0$ to suppress such a mixing between the Higgs and NG modes.\cite{Varma2002,PekkerVarma2015}
In other words, the Higgs mode is to be protected by the particle-hole symmetry.
However, this argument for the stability of the Higgs mode is not needed 
in the case of charged superconductors (although the particle-hole
symmetry is a good symmetry near the Fermi surface in superconductors),
since {\it the NG field decouples from the Higgs field [in Eq.~(\ref{GL higgs})] due to the Anderson-Higgs mechanism} as stated above.

\begin{figure}[t]
\includegraphics[width=4cm]{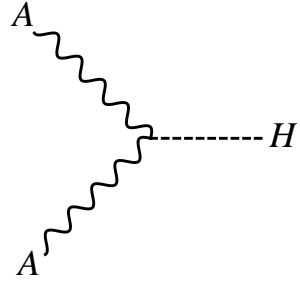}
\caption{
The Feynman diagram for the interaction vertex that connects
the Higgs field $H$ (dashed line) and electromagnetic field $\bm A$ (wavy lines).
}
\label{A2h diagram}
\end{figure}

Equation (\ref{GL higgs}) suggests that the interaction between the Higgs and gauge fields is given by 
$\phi H$, $\phi^2 H$ and $\bm A^2 H$. The linear coupling $\phi H$ is suppressed in superconductors
due to the inherent particle-hole symmetry ($d\sim 0$).
The leading interaction is the second-order process $\phi^2 H$ and $\bm A^2 H$,
the latter of which, e.g., is represented by a Feynman diagram shown in Fig.~\ref{A2h diagram}.
%The Higgs-gauge coupling necessarily becomes nonlinear with respect to $\phi$ and $\bm A$,
The nonlinear Higgs-gauge coupling implies that $H$ describes a scalar boson having no electric charge.
%and hence the linear coupling term must vanish.
These nonlinear couplings ($\phi^2 H$ and $\bm A^2 H$) have indeed been used in the discovery of the Higgs particle 
at the LHC experiment\cite{ATLAS2012,CMS2012}
(where $\bm A$ corresponds to the vector bosons $W$ or $Z$).

%\begin{align*}
%-c\partial_t^2 \Psi
%&=
%a\Psi+b|\Psi|^2\Psi-\frac{1}{2m^\ast}(\nabla-ie^\ast\bm A)^2\Psi
%\end{align*}

From Eq.~(\ref{GL higgs}) (with $d=0$), we can derive the equation of motion for the Higgs field,
\begin{align}
\left(c\partial_t^2-\frac{1}{2m^\ast}\nabla^2\right) H
&=
2aH+e^\ast{}^2\Psi_0\left(c\phi^2-\frac{1}{2m^\ast}\bm A^2\right),
\label{Higgs equation of motion}
\end{align}
which is  ``relativistic'' (with an emergent Lorentz symmetry),\cite{PekkerVarma2015}
meaning that the first time-derivative is absent even though we started from the non-relativistic GL
Lagrangian (\ref{GL Lagrangian}).
Let us first look at the case of $\phi=\bm A=0$. By putting
$H(\bm r,t)\sim e^{i\bm q\cdot \bm r-i\omega t}$, we obtain the dispersion relation
for the Higgs mode,
\begin{align}
\omega(\bm q)^2
&=
-\frac{2a}{c}+\frac{q^2}{2m^\ast c}
=
\omega_H^2+\frac{q^2}{2m^\ast c},
\end{align}
where the mode is a gapped (massive) excitation with a characteristic frequency (mass) 
\begin{align}
\omega_H=\sqrt{-\frac{2a}{c}}.
\end{align}
From this, one can see that $\omega_H\propto (T_c-T)^{1/2}\propto \Delta$.
In fact, the microscopic calculation\cite{Anderson1958,LittlewoodVarma1981a}
shows that $\omega_H=2\Delta$, which exactly coincides with
the lowest energy necessary to create a pair of Bogoliubov quasiparticles.
Using the microscopic result\cite{Gorkov1959}
of $b(\Psi_0/\Delta)^2=3/(4\epsilon_F)$ (with $\epsilon_F$ the Fermi energy),
we have $c=-2a/(2\Delta)^2=2b\Psi_0^2/(2\Delta)^2=3/(8\epsilon_F)$. With this and $m^\ast=2m$,
we reproduce the well-known relation,\cite{LittlewoodVarma1981a}
\begin{align}
\omega(\bm q)^2
&=
(2\Delta)^2+\frac{1}{3}v_F^2 q^2,
\end{align}
where $v_F=\sqrt{2\epsilon_F/m}$ is the Fermi velocity.
From Eq.~(\ref{GL expanded}), it is obvious that the dispersion for the NG mode (or Bogoliubov mode)
shares the same form $\omega(\bm q)^2=q^2/(2m^\ast c)=v_F^2q^2/3$ with the Higgs mode besides the mass term.
This agrees with the previously known result.\cite{BogoliubovBook}

Next, we turn to a situation where the system is driven by a continuous and homogeneous ac electric field $\bm A(t)=\bm A e^{-i\Omega t}$.
The problem becomes equivalent to a forced oscillation of a harmonic oscillator,
and the solution for Eq.~(\ref{Higgs equation of motion}) is given by
\begin{align}
H(t)
&=
\frac{1}{(2\Omega)^2-\omega_H^2}\frac{e^\ast{}^2\Psi_0A^2}{2m^\ast c} e^{-2i\Omega t}.
\label{GL result}
\end{align}
This captures the fundamental aspect of the resonance phenomenon discussed in the paper.
Due to the nonlinear coupling to the electric field, the elementary frequency of the oscillation of the Higgs field
is $2\Omega$ (rather than $\Omega$). When $2\Omega$ matches with the eigenfrequency of the Higgs field $\omega_H$,
the resonance occurs and the oscillation amplitude diverges as $(2\Omega-\omega_H)^{-1}$. 
From a microscopic point of view, this phenomenon can be understood  as a resonant
precession of Anderson pseudospins as we shall discuss in Sec.~\ref{time dependent BdG}.

The current $\bm j=\partial\mathcal L/\partial\bm A$ is expressed as
\begin{align*}
\bm j
&=
-\frac{ie^\ast}{2m^\ast}[\Psi^\dagger \nabla \Psi-(\nabla\Psi^\dagger)\Psi]
-\frac{e^\ast{}^2}{m^\ast}\bm A\Psi^\dagger \Psi.
\end{align*}
Expanding $\Psi$ around $\Psi_0$, we obtain the leading nonlinear current response
against $\bm A$,
\begin{align*}
\bm j_{\rm NL}(t)
&=
-\frac{2e^\ast{}^2\Psi_0}{m^\ast}\bm A(t) H(t).
\end{align*}
This takes the form of a London equation, where the current is proportional to $\bm A(t)$.
Remarkably, the nonlinear current is also {\it proportional to the Higgs field} $H(t)$,
so that the current can, and does indeed, sensitively reflect the temporal change of the Higgs field.
Since $\bm A(t)$ oscillates with frequency $\Omega$, while $H(t)$ oscillates with $2\Omega$,
the current [$\propto \bm A(t) H(t)$] ends up with oscillating with
frequency $3\Omega$. This implies that 
a {\it giant third harmonic generation} (THG) is induced near the resonance ($2\Omega\sim \omega_H$)
with the Higgs mode.

\begin{figure}[t]
\includegraphics[width=7cm]{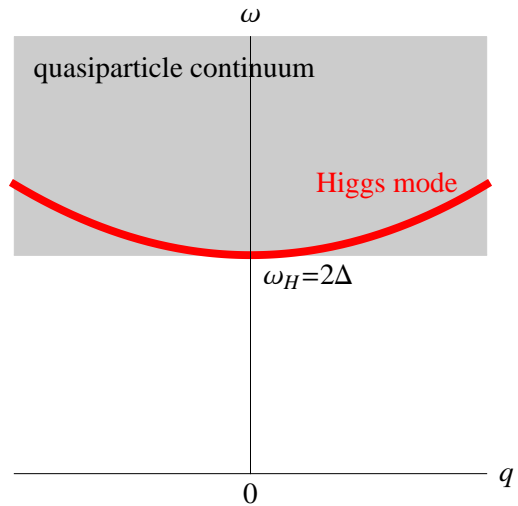}
\caption{
(Color online)
Schematic excitation spectrum of $s$-wave superconductors.
The red curve is the collective Higgs mode, while the shaded region represents
the quasiparticle excitation continuum.
}
\label{energy spectrum}
\end{figure}

So far, we have discussed the Higgs mode and its resonance with electromagnetic waves 
based on the time-dependent GL theory (\ref{GL Lagrangian}). 
Apart from the fact that the characteristic frequency $\omega_H$ of the Higgs mode
cannot be determined within GL theory, 
the problem of GL theory is that 
it does not take account of relaxations of the Higgs mode into quasiparticles.
As shown in Fig.~\ref{energy spectrum},
the Higgs mode is degenerate with the lower bound of the quasiparticle excitation continuum 
($\omega_H=2\Delta$). The coincidence of the two energies is known as the Nambu relation.\cite{Nambu1985,VolovikZubkov}
Since the Higgs mode lies at the same energy scale as the quasiparticle excitations,
it can easily decay into individual quasiparticles.
Furthermore, at low temperatures the relaxation time of quasiparticles becomes much longer
than the time scale of the order-parameter variation in clean superconductors.
As a result, one cannot neglect quasiparticle excitations, and
the dynamics of the order parameter is necessarily entangled with
those of quasiparticles. 
%One may phenomenologically include a damping term such as $\gamma\Psi^\dagger \partial_t\Psi$ in Eq.(\ref{GL Lagrangian}).
%However, it just adds an imaginary part of the coefficient $d$, so that the effect of the damping is erased by the Anderson-Higgs mechanism,
%and the result remains the same as Eq.~(\ref{GL higgs}).
The low-energy effective theory of the Higgs mode may not be expressed only in terms of $\Psi$,
but may involve fermionic degrees of freedom. The crucial questions that arise are
(i) whether the Higgs resonance discussed here
would survive or not after we take account of the relaxation
to quasiparticles (pair-breaking process),
and (ii) if it would survive, then how one can distinguish the collective Higgs mode
from individual quasiparticle excitations, both of which are energetically degenerate.
These motivate us to move on to
the underlying microscopic theory in the subsequent sections.

\section{Microscopic theory for Anderson pseudospin resonance}
\label{time dependent BdG}

Having identified the necessity of going beyond GL theory,
we start from the pairing Hamiltonian
for an $s$-wave superconductor coupled to a dynamical electric field,
\begin{align}
H_{\rm pair}=\sum_{\bm k, \sigma}\epsilon_{\bm k-e\bm A(t)}c_{\bm k\sigma}^\dagger c_{\bm k\sigma}
-U\sum_{\bm k, \bm p} c_{\bm k\uparrow}^\dagger  c_{-\bm k\downarrow}^\dagger c_{-\bm p\downarrow}c_{\bm p\uparrow},
\label{pairing Hamiltonian}
\end{align}
where $\epsilon_{\bm k}$ is the band dispersion measured from Fermi energy $\epsilon_F$, 
$e$ the elementary charge,
$\bm A(t)=\bm A \sin\Omega t$ the vector potential for the ac electric field $\bm E(t)$ introduced by Peierls substitution (in the temporal gauge),
$c_{\bm k\sigma}^{\dagger}$ the creation operator for electrons,
and $-U(<0)$ is the attractive pairing interaction. We consider a superconducting thin film,
into which the electric field can penetrate.
%Our focus is on the superconducting state during irradiation of the laser field, for which we take an ac electric field, i.e., $\bm A(t)=\bm A \sin\Omega t$.
For $H_{\rm pair}$ (\ref{pairing Hamiltonian}), the BCS mean-field description becomes exact. 
%We rewrite
%$H_{\rm pair}=\sum_{\bm k}\Psi_{\bm k}^\dagger \mathcal H_{\rm BdG} \Psi_{\bm k}$ 
%with the Bogoliubov-de Gennes (BdG) Hamiltonian
%\begin{align}
%H_{BCS}&=\sum_{\bm k}
%\Psi_{\bm k}^\dagger
%\mathcal H_{\rm BdG}
%\Psi_{\bm k},
%\;\;
%\mathcal H_{\rm BdG}
%=
%\begin{pmatrix}
%\epsilon_{\bm k-e\bm A(t)} & -\Delta(t)^\ast \\
%-\Delta(t) & -\epsilon_{\bm k+e\bm A(t)}
%\end{pmatrix},
%\label{BCS Hamiltonian}
%\end{align}
We define the superconducting gap function, 
%as an expectation value $(\langle \rangle)$, 
\begin{align}
\Delta=\Delta'+i\Delta''=U\sum_{\bm k}\langle c_{\bm k\uparrow}^\dagger c_{-\bm k\downarrow}^\dagger \rangle,
\label{gap function}
\end{align}
that serves as the order parameter in the BCS theory.
We can replace the momentum sum with an integral $D(\epsilon_F) \int_{-\omega_D}^{\omega_D} d\epsilon$
with $D(\epsilon_F)$ the density of states at the Fermi energy and $\omega_D$ 
the energy cut off (e.g., the Debye frequency of the bosonic pairing glue such as phonons).
The interaction strength is characterized by a dimensionless $\lambda=UD(\epsilon_F)$.
In the following we set $\hslash=1$, and use $\omega_D$ as the unit of energy.

\begin{figure}[b]
\includegraphics[width=6cm]{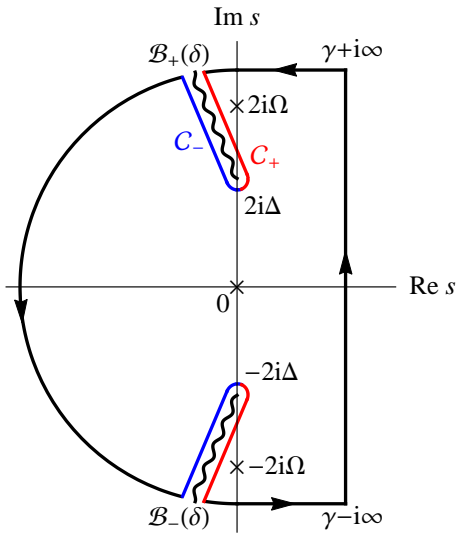}
\caption{(Color online) The integral contour (solid closed curve) that we take on the complex plane to evaluate the integral (\ref{Bromwich}).
Crosses and wavy lines represent poles and branch cuts, respectively.}
%The green contour $\widetilde{\mathcal C}_-$ is the steepest descent curve used in the saddle-point method.}
\label{fig:contour}
\end{figure}

%To describe the dynamics, we use 
Anderson's pseudospin \cite{Anderson1958} is defined by
\begin{align}
\bm\sigma_{\bm k}&=\frac{1}{2}\Psi_{\bm k}^\dagger \cdot \bm \tau \cdot \Psi_{\bm k},
\end{align}
where $\Psi_{\bm k}=(c_{\bm k\uparrow}, c_{-\bm k\downarrow}^\dagger)^t$ is the Nambu spinor,
and $\bm\tau=(\tau^x,\tau^y,\tau^z)$ are the Pauli matrices. The pseudospin satisfies the usual commutation relations for angular momentum,
$[\sigma_{\bm k}^j, \sigma_{\bm k}^k]=i\varepsilon_{jkl}\sigma_{\bm k}^l$.
%The $z$ component $\sigma_{\bm k}^z=(n_{\bm k\uparrow}+n_{\bm k\downarrow}-1)/2$ represents the occupation of particles, while
%$x$ and $y$ components represent $U(1)$ gauge symmetry breaking.
With this, the pairing Hamiltonian (\ref{pairing Hamiltonian}) is 
recast in a form, 
\begin{align}
H_{\rm pair}&=2\sum_{\bm k} \bm b_{\bm k}\cdot \bm \sigma_{\bm k},
\end{align}
which can be regarded as a spin system in an effective magnetic field, 
%defined by
\begin{align}
\bm b_{\bm k}&=\left(-\Delta',-\Delta'',\frac{\epsilon_{\bm k-e\bm A(t)}+\epsilon_{\bm k+e\bm A(t)}}{2}\right).
\end{align}
The $z$ component of $\bm b_{\bm k}$ represents the light-matter coupling involving
contributions from both the particle and hole sectors.
Since $\bm b_{\bm k}$ is a function even in $\bm A(t)$ if the system is parity symmetric ($\epsilon_{-\bm k}=\epsilon_{\bm k}$), 
%the vector potential appears as an average of $\epsilon_{\bm k-e\bm A(t)}$ and $\epsilon_{\bm k+e\bm A(t)}$ in ,
we can readily recognize that the linear coupling vanishes, 
so that the leading effect of the electric field starts from 
$O(\bm A(t)^2)$. 
The self-consistency condition (\ref{gap function}) reads
\begin{align}
\Delta=U\sum_{\bm k}(\sigma_{\bm k}^x+i\sigma_{\bm k}^y)
\end{align}
in the pseudospin notation. 
While the dynamics cannot be described by the conventional GL equation,
which would be valid only when the time scale of the order-parameter motion is
much longer than that of quasiparticle relaxations,
in the present formalism the time evolution is determined by 
%the BdG equation,
%$i\partial_t\Psi_{\bm k}(t)=\mathcal H_{\rm BdG}(t)\Psi_{\bm k}(t)$,
%or equivalently, by 
a Bloch equation for the pseudospins,
%that describes Anderson's pseudospin precession
\cite{VolkovKogan1973,BarankovLevitovSpivak2004,Yuzbashyan2005}
\begin{align}
\partial_t\bm \sigma_{\bm k}
&=
i[H_{\rm pair},\bm\sigma_{\bm k}]
=
2\bm b_{\bm k}\times \bm \sigma_{\bm k}.
\label{Bloch equation}
\end{align}
Anderson pseudospins have been recently used to analyze the dynamics
of charge fluctuations in a time-resolved Raman experiment for high-$T_c$ cuprates.
\cite{Mansart2013,Lorenzana2013}

\begin{figure}
\includegraphics[width=8cm]{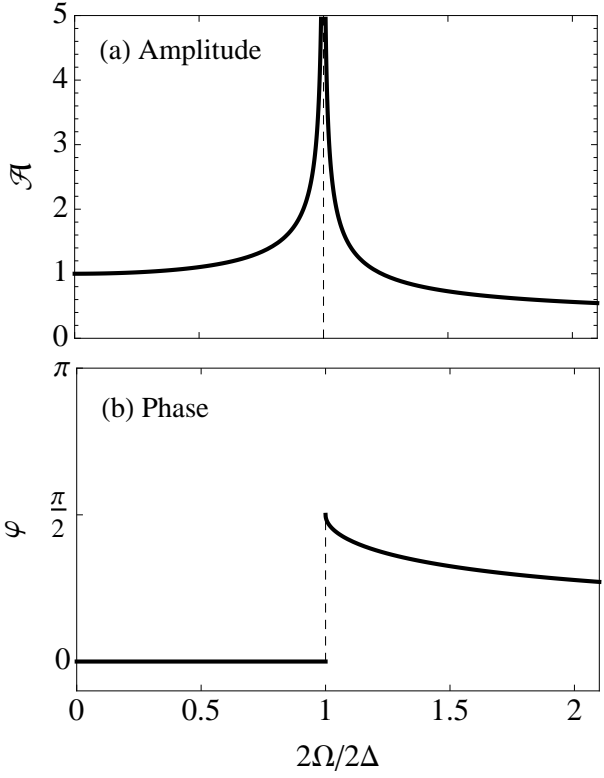}
\caption{(a) The amplitude $\mathscr A$ and (b) phase shift $\varphi$ of the $2\Omega$ oscillation of the superconducting order parameter 
$\delta\Delta(t)$ [Eq.~(\ref{asymptotic})] against $2\Omega/2\Delta$.}
%Both of $\mathscr A$ and $\varphi$ are functions of $(2\Omega/2\Delta)$.}
\label{fig:BdG}
\end{figure}

We can analytically solve Eq.~(\ref{Bloch equation}) up to the leading (second) order in $\bm A(t)$. This is achieved by
linearizing Eq.~(\ref{Bloch equation}) with the time-independent and time-dependent parts separated as 
$\bm\sigma_{\bm k}(t)=\bm\sigma_{\bm k}(0)+\delta\bm\sigma_{\bm k}(t)$ and $\Delta(t)=\Delta+\delta\Delta(t)$.
We assume that the initial state is superconducting at zero temperature. 
The initial $\Delta$ may be taken to be real positive without loss of 
generality. Thus the initial condition 
reads 
$\sigma_{\bm k}^x(0)=\Delta/\omega_{\bm k}$ and $\sigma_{\bm k}^z(0)=-\epsilon_{\bm k}/\omega_{\bm k}$
with $\omega_{\bm k}=2(\epsilon_{\bm k}^2+\Delta^2)^{1/2}$. The linearized equations of motion are
\begin{align}
&\partial_t \delta\sigma_{\bm k}^x(t)
=
-2\epsilon_{\bm k}\delta\sigma_{\bm k}^y(t),
\label{BdG1}
\\
&\partial_t \delta\sigma_{\bm k}^y(t)
=
2\epsilon_{\bm k}\,\delta\sigma_{\bm k}^x(t)+2\Delta\, \delta\sigma_{\bm k}^z(t)
\nonumber
\\
&\quad
+\frac{1}{\omega_{\bm k}}
\bigg[e^2\Delta \sum_{ij}\partial_{k_i}\partial_{k_j}\epsilon_{\bm k}A_i(t)A_j(t)
-2\epsilon_{\bm k}\delta\Delta(t)\bigg],
\label{BdG2}
\\
&\partial_t \delta\sigma_{\bm k}^z(t)
=
-2\Delta\, \delta\sigma_{\bm k}^y(t).
\label{BdG3}
\end{align}
Note that $\partial_t(\Delta\delta\sigma_{\bm k}^x-\epsilon_{\bm k}\delta\sigma_{\bm k}^z)=0$.
From this, along with the initial condition $\delta\bm \sigma_{\bm k}(0)=0$, it turns out that 
the relation $\Delta\delta\sigma_{\bm k}^x=\epsilon_{\bm k}\delta\sigma_{\bm k}^z$ holds all the time,
which helps us to reduce the number of the equations.
%\begin{align*}
%\left(\frac{\partial^2}{\partial t^2}+\omega_{\bm k}^2\right)\frac{\delta\sigma_{\bm k}^x(t)}{\Delta}
%=
%\frac{4\epsilon_{\bm k}^2}{\omega_{\bm k}}
%\left(\frac{\delta\Delta(t)}{\Delta}
%-\sum_{ij}\frac{1}{2\epsilon_{\bm k}}\frac{\partial^2\epsilon_{\bm k}}{\partial k_i\partial k_j}A_i(t)A_j(t)\right)
%\end{align*}
%with $\delta\Delta(t)=U\sum_{\bm k}\delta\sigma_{\bm k}^x$.

We solve the equations by a Laplace transformation, $\mathcal L[\delta\Delta(t)](s)=:\delta\Delta(s)$, etc.
Let us call the direction of the electric field $x$. Then we have
$\sum_{ij}\partial_{k_i}\partial_{k_j}\epsilon_{\bm k} A_i(t)A_j(t)=\partial_{k_x}^2\epsilon_{\bm k}|\bm A(t)|^2$.
When the crystallographic directions are equivalent, 
we have $\partial_{k_x}^2\epsilon_{\bm k}\to d^{-1}\nabla_{\bm k}^2\epsilon_{\bm k}$
with $d$ the spatial dimension. 
%After several manipulations, we obtain
%\begin{align*}
%\sum_{\bm k}\frac{s^2+4\Delta^2}{\omega_{\bm k}(s^2+\omega_{\bm k}^2)}\frac{\delta\Delta(s)}{e^2A^2\Delta}
%&=
%-\frac{\Omega^2}{s(s^2+4\Omega^2)}\sum_{\bm k}\frac{4\epsilon_{\bm k}d^{-1}\nabla_{\bm k}^2\epsilon_{\bm k}}{\omega_{\bm k}(s^2+\omega_{\bm k}^2)}.
%\end{align*}
%To proceed further, we need an explicit form for $\epsilon_{\bm k}$.
If the band dispersion is isotropic with $\epsilon_{\bm k}=\epsilon(|\bm k|)$,
we expand 
$\epsilon_{\bm k}$ around the Fermi wave number $k_F$ as $\epsilon_{\bm k}=\sum_{n=1}^\infty c_n (|\bm k|-k_F)^n$.
With this, we can define a series expansion,
\begin{align}
d^{-1}\nabla_{\bm k}^2\epsilon_{\bm k}=\alpha_0+\alpha_1\epsilon_{\bm k}+\alpha_2\epsilon_{\bm k}^2+\cdots,
\end{align}
where $\alpha_0=2c_2d^{-1}+c_1(1-d^{-1})k_F^{-1}, \alpha_1=c_1^{-1}[6c_3d^{-1}+(1-d^{-1})(2c_2k_F^{-1}-c_1k_F^{-2})]$, etc.,
with each coefficient $\alpha_n$ $\sim O(\epsilon_F^{1-n})$.
%Since the terms odd-order in $\epsilon_{\bm k}$ vanish in the momentum integral, 
Since the $\alpha_0$ term just gives a trivial phase $\exp(i\alpha_0e^2\int_0^t dt'\, A(t')^2)$ to $\Delta(t)$, which can be gauged out,
the $\alpha_1$ term provides the leading contribution around the Fermi surface 
(with $\omega_D\ll \epsilon_F$). 
%The rest is of higher orders in $(|\bm k|-k_F)$.
For anisotropic band structures, the same expansion is still sometimes possible.
For instance, the $d$-dimensional cubic lattice 
(with cosine bands $\epsilon_{\bm k}=-2\sum_i \cos k_i-\epsilon_F$) has $d^{-1}\nabla_{\bm k}^2\epsilon_{\bm k}=\alpha_0+\alpha_1\epsilon_{\bm k}$ 
with $\alpha_0=-\epsilon_F d^{-1}$ and $\alpha_1=-d^{-1}$.

Thus, in most cases of our interest, we arrive at
%\begin{align}
%\delta\sigma_{\bm k}^x(s)
%&=
%\frac{4\epsilon_{\bm k}}{\omega_{\bm k}(s^2+\omega_{\bm k}^2)}
%\left(\epsilon_{\bm k}\delta\Delta(s)-\frac{\Delta \Omega^2}{s(s^2+4\Omega^2)}\sum_{ij}\frac{\partial^2\epsilon_{\bm k}}{\partial k_i\partial k_j}A_iA_j\right)
%\end{align}
\begin{align}
\frac{\delta\Delta(s)}{\alpha_1 e^2A^2\Delta}
&=
\frac{\Omega^2}{s(s^2+4\Omega^2)}
\left[1-\frac{1}{\lambda (s^2+4\Delta^2)F(s)}\right],
\end{align}
where $A=|\bm A|$ and
%$\lambda=UD(\epsilon_F)$ and 
\begin{align}
F(s)&=\int_{-\infty}^{\infty} d\epsilon\, \frac{1}{2\sqrt{\epsilon^2+\Delta^2}(s^2+4\epsilon^2+4\Delta^2)}
\nonumber
\\
&=\frac{1}{s\sqrt{s^2+4\Delta^2}}\sinh^{-1}\left(\frac{s}{2\Delta}\right).
\end{align}
In the above, we have replaced the range of integration from $\int_{-\omega_D}^{\omega_D}$ to $\int_{-\infty}^{\infty}$,
which is allowed in the BCS regime ($\omega_D\gg \Delta$).
$F(s)$ can be analytically continued on the complex plane, where 
branch cuts $\mathcal B_\pm(\delta)=\{\pm 2i\Delta\pm ire^{\pm i\delta}|r\in [0,\infty)\}$ with 
$\delta$ small but nonzero are introduced (Fig.~\ref{fig:contour}).

To obtain $\delta\Delta(t)$ with an inverse Laplace transformation, we need to evaluate a Bromwich integral,
\begin{align}
I(t)
&=
\frac{1}{2\pi i}\int_{\gamma-i\infty}^{\gamma+i\infty}ds\,
e^{st} \frac{\Omega^2}{s(s^2+4\Omega^2)(s^2+4\Delta^2)F(s)},
\label{Bromwich}
\end{align}
where $\gamma\in\mathbb R$ is taken to be larger than any of the real parts of the poles in the integrand.  
%While this expression may seem complicated, essential point is the following. 
There are three first-order poles at $s=0, \pm 2i\Omega$ and two branching points at $s=\pm 2i\Delta$ (Fig.~\ref{fig:contour}) in the integrand, where 
$s=\pm 2i\Delta$ corresponds to the Higgs amplitude mode, while $s=\pm 2i\Omega$ to the forced precession of the Anderson pseudospins driven by the electric field.
As one changes $\Omega$, the poles merge with the branching points at $2\Omega=2\Delta$, 
which causes a {\it resonance} between the forced pseudospin precession and the Higgs mode.

To make it more explicit, 
we evaluate the integral (\ref{Bromwich}) by taking the contour as depicted in Fig.~\ref{fig:contour}, which surrounds the three poles
but avoids the branch cuts. This kind of contour is often used to calculate similar integrals (see, e.g., Ref.~\cite{VolkovKogan1973}).
We take $\delta>0$ so that the contours along the branch cuts $\mathcal B_\pm(\delta)$ 
do not touch the poles $s=\pm 2i\Omega$
when $\Omega>\Delta$.
Since the contributions from infinity vanish,
we are left with the residues of the poles and the line integrals ($\mathcal C_\pm$ in Fig.~\ref{fig:contour} and their Hermitian conjugates) along the branch cuts.
The asymptotic behavior of the integrals $\mathcal C_\pm$ for $t\to\infty$ is evaluated by the saddle-point method. 
%The integrals $\mathcal C_+$ and $\mathcal C_-$ are related with each 
%other with flipped signs of $\sqrt{s^2+4\Delta^2}$.
%As a result, we obtain
%\begin{align}
%&I(t)=
%\frac{1}{16\Delta^2F(0)}-{\rm Re} \frac{e^{2i\Omega t}}{32(\Delta^2-\Omega^2)F(2i\Omega)}
%\nonumber
%\\
%&
%+{\rm Re}\int_{\mathcal C_-} ds\, e^{st}\frac{\Omega^2}{s^2(s^2+4\Omega^2)(s^2+4\Delta^2)^{\frac{3}{2}}F_+(s)F_-(s)},
%\end{align}
%where we have used $F_+(s)-F_-(s)=-i\pi/(s\sqrt{s^2+4\Delta^2})$ for $s\in \mathcal C_-$.
%For the integral $\mathcal C_-$, we replace the integral variable $s=\Delta(z-z^{-1})$,
%and deform the contour without crossing singularities 
%into the steepest descent curve. 
%that corresponds to $\widetilde{\mathcal C}_-$ in Fig.~\ref{fig:contour}.
%The dominant contribution to the asymptotic behavior arises near $z=i$.
%The saddle-point method gives the asymptotic form of the third term in Eq.~() as
%$\sim\frac{1}{2\lambda}\frac{1}{\pi^{3/2}}\frac{\Omega^2}{\Delta^2-\Omega^2}\frac{1}{\sqrt{\Delta t}}\cos\left(2\Delta t+\frac{\pi}{4}\right)$.
Finally we end up with long-time asymptotic forms of the order parameter,
\begin{widetext}
\begin{align}
&\frac{\delta\Delta(t)}{\alpha_1 e^2A^2\Delta}
\sim
\frac{1}{4\lambda}\left[
\frac{2}{\pi^{3/2}}\frac{\Omega^2}{\Omega^2-\Delta^2}\frac{1}{\sqrt{\Delta t}}\cos\left(2\Delta t+\frac{\pi}{4}\right)
-1\right]
+\frac{1-\cos 2\Omega t}{4}
+\frac{1}{4\lambda}\times
\begin{cases}
\displaystyle
\frac{\Omega}{\sqrt{\Delta^2-\Omega^2}}\frac{\cos 2\Omega t}{\sin^{-1}\left(\frac{\Omega}{\Delta}\right)} & \Omega<\Delta
\vspace{.1cm}
\\
\displaystyle
\frac{\Omega}{\sqrt{\Omega^2-\Delta^2}}
\frac{\cos (2\Omega t-\varphi)}{\sqrt{[\cosh^{-1}\left(\frac{\Omega}{\Delta}\right)]^2+\left(\frac{\pi}{2}\right)^2}}
& \Omega>\Delta
\end{cases},
\label{asymptotic}
\end{align}
\end{widetext}
where $\varphi$ is the phase shift given by
\begin{align}
\varphi=\tan^{-1}\left(\frac{\pi/2}{\cosh^{-1}\left(\frac{\Omega}{\Delta}\right)}\right).
\end{align}

The first term in Eq.~(\ref{asymptotic}) can be interpreted as the Higgs amplitude mode
induced by an effective change of the interaction parameter due to the ac field,
$U\to U_{\rm eff}=(1-\frac{1}{2}\alpha_1 e^2A^2)U$. \cite{TsujiOkaAoki2009}
Indeed, it approaches the result for the interaction-quench problem \cite{BarankovLevitov2006,YuzbashyanDzero2006} in the limit of $\Omega\to\infty$.
The Higgs mode is amplified by the ac electric field around $2\Omega=2\Delta$.
The term decays algebraically as $t^{-1/2}$,\cite{VolkovKogan1973} which suggests that the Higgs mode effectively has an infinite lifetime within the BCS approximation.

In the long-time limit, the constant term and the term oscillating with frequency $2\Omega$ survive.
The constant term in $\delta\Delta(t)$ is proportional to $\alpha_1(1-\lambda^{-1})$, 
which implies, intriguingly, that 
we can attain an {\it amplification} of superconductivity 
on time average when this term is positive. 
The $2\Omega$ oscillation term represents the APR.
If we write the last term in Eq.~(\ref{asymptotic}) as
$\frac{1}{4\lambda}\mathscr{A}\cos(2\Omega t-\varphi)$, the amplitude $\mathscr A$ and the phase shift $\varphi$
are universal functions that depend only on the ratio $2\Omega/2\Delta$ (Fig.~\ref{fig:BdG}).
The amplitude $\mathscr A$ diverges as $|2\Omega-2\Delta|^{-1/2}$ at $2\Omega=2\Delta$ (resonance condition).
It clearly differs from the result of the time-dependent GL (\ref{GL result}), $|2\Omega-2\Delta|^{-1}$.
The reduction of the power from $1$ to $1/2$ signifies that the Higgs mode is a bit less stable, where each pseudospin precession gradually dephases.
Physically we can interpret  this as coming from Landau damping; that is, the collective mode decays into individual quasiparticle excitations
even in the collision-less equation (\ref{Bloch equation}).
An anomaly is also found in the phase shift $\varphi$: for $2\Omega<2\Delta$, $\varphi$ is locked to zero, i.e., the $2\Omega$ oscillation of the order parameter is in-phase 
with $\bm E(t)^2$. As soon as $2\Omega$ exceeds $2\Delta$, the $\varphi$ discontinuously jumps to $\pi/2$ and starts to drift (Fig.~\ref{fig:BdG}).
Along with the order-parameter oscillation, 
the pseudospin itself continues to precess around the axis parallel to $\bm \sigma_{\bm k}(0)$, 
with two modes of frequencies $\omega_{\bm k}$ and $2\Omega$ surviving in $t\to\infty$ (the former of which dephases).
By numerically simulating Eq.~(\ref{Bloch equation}), we also 
confirmed that APR generally occurs for finite-temperature initial states 
and for pulsed electric fields that contain large enough number of oscillation cycles.
%Therefore, APR is a quite general phenomenon.
%The fact that the poles of the Higgs mode lies on the imaginary axis suggests that the Higgs mode has an infinite lifetime,
%which is consistent with the observation that the amplitude oscillation decays algebraically as $t^{-1/2}$.

APR appears in various physical quantities. What is readily 
accessible experimentally is
the electric current,
\begin{align}
\bm j=e\sum_{\bm k, \sigma} \bm v_{\bm k-e\bm A(t)}c_{\bm k\sigma}^\dagger c_{\bm k\sigma}
\end{align}
($\bm v_{\bm k}=\nabla_{\bm k} \epsilon_{\bm k}$ is the group velocity).
%and $n_{\bm k\sigma}=c_{\bm k\sigma}^\dagger c_{\bm k\sigma}$]. 
The current is expressed in the pseudospin notation
as $\bm j=e\sum_{\bm k} [\bm v_{\bm k-e\bm A(t)}-\bm v_{\bm k+e\bm A(t)}]\sigma_{\bm k}^z
+\frac{e}{2}\sum_{\bm k\sigma}[\bm v_{\bm k-e\bm A(t)}+\bm v_{\bm k+e\bm A(t)}]c_{\bm k\sigma}^\dagger c_{\bm k\sigma}$. 
If we expand it in $\bm A(t)$, 
the linear response is given by $\bm j^{(1)}=-2\alpha_1 e^2 \bm A(t)\sum_{\bm k} \epsilon_{\bm k}\sigma_{\bm k}^z(0)
+e\sum_{\bm k\sigma}v_{\bm k}c_{\bm k\sigma}^\dagger c_{\bm k\sigma}$,
which is irrelevant to APR. 
%(hence it is not resonantly enhanced).
In fact, the linear-response optical conductivity does not show any divergence for $\Omega\neq 0$. \cite{Zimmermann1991}
The leading term that reflects the change of the order parameter is the third-harmonic generation (THG),
\begin{align}
\bm j^{(3)}(t)=-2\alpha_1 e^2\Delta U^{-1}\delta\Delta(t)\bm A(t),
\label{j3 Delta}
\end{align}
where we have used $\epsilon_{\bm k}\delta\sigma_{\bm k}^z=\Delta\delta\sigma_{\bm k}^x$.
%where we have used the relation $\epsilon_{\bm k}\delta\sigma_{\bm k}^z=\Delta \delta\sigma_{\bm k}^x$.
%It is proportional to $\delta\Delta(t)$.
%Since $\delta\Delta(t)$ is of second order in $\bm A(t)$, $\bm j^{(3)}$ is of third order in total. 
%We note Eq.(\ref{THG}) takes the form of London equation. In this sense, one can say that THG arises as part of the supercurrent.
The consequence is remarkable: although the frequency $\Omega$ is below the energy gap, 
% so that quasiparticle excitation is suppressed,
%(especially the linear-response optical conductivity does not show any divergence for $\Omega\neq 0$),
we do obtain the colossal nonlinear response due to divergence of $\delta\Delta(t)$. It may be used as an efficient THz harmonic emitter.

\section{Response function for Anderson pseudospin resonance}

To reinforce our picture for APR phenomenon,
we can approach it from an alternative, diagrammatic point of view.
%In this section, we look at the APR phenomenon from an alternative diagrammatic point of view.
This allows one to decompose the THG susceptibility into the bare and vertex-correction
diagrams, each of which contains individual and collective excitations, respectively.
Thus we can unambiguously distinguish the Higgs mode from quasiparticle excitations that are degenerate
at the superconducting gap energy.
To this end, we take the Nambu-Gor'kov Green's function defined by
\begin{align*}
\hat G_{\bm k}(t,t')
&=
\begin{pmatrix}
-i\langle \mathcal T c_{\bm k\uparrow}(t)c_{\bm k\uparrow}^\dagger(t')\rangle & -i\langle \mathcal T c_{\bm k\uparrow}(t)c_{-\bm k\downarrow}(t')\rangle \\
-i\langle \mathcal T c_{-\bm k\downarrow}^\dagger(t) c_{\bm k\uparrow}^\dagger(t')\rangle & -i\langle \mathcal T c_{-\bm k\downarrow}^\dagger(t)c_{-\bm k\downarrow}(t')\rangle
\end{pmatrix},
\end{align*}
where $\mathcal T$ represents the time ordering.
With this, the gap function is expressed as
\begin{align}
\Delta(t)
&=
-\frac{i}{2}U\sum_{\bm k} {\rm Tr}\left[\tau_1\hat G_{\bm k}^<(t,t)\right],
\label{Delta Green function}
\end{align}
where $\hat G_{\bm k}^<$ is the lesser Green's function,
and $\Delta(t)$ is assumed to be real.

\begin{figure}[t]
\includegraphics[width=8cm]{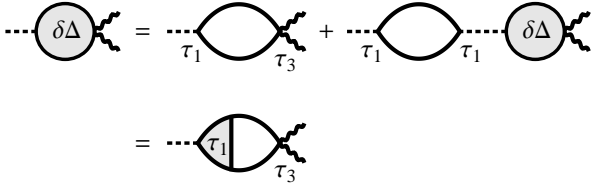}
\caption{
The diagrammatic representation of the self-consistent Eq.~(\ref{self consistent eq for Delta}) for $\delta\Delta$
and its relation to the $\tau_1$ vertex. The wavy and dashed lines represent the gauge field $\bm A$ and the interaction vertices,
respectively.
}
\label{delta feynman diagram}
\end{figure}

Now we take variations of both sides of Eq.~(\ref{Delta Green function})
with respect to the external field $\bm A(t)$. In this section, we consider the monochromatic wave $\bm A(t)=\bm A e^{-i\Omega t}$.
Since the leading change of the order parameter $\delta\Delta(t)=\delta\Delta e^{-2i\Omega t}$
is the second order in $\bm A(t)$, we have
\begin{align}
\delta\Delta
&=
\delta_A^2\Delta
=
-\frac{i}{2}U\sum_{\bm k} {\rm Tr}\left[\tau_1\delta_A^2\hat G_{\bm k}^<\right],
\end{align}
where $\delta_A$ represents the functional derivative with respect to $\bm A$.
%and $<$ denotes the lesser component of Green's function.
In the BCS theory, the Nambu-Gor'kov Green's function is given by
the Dyson equation $\hat G_{\bm k}=(i\partial_t-\hat \xi_{\bm k}+\Delta\tau_1)^{-1}$
with $\hat \xi_{\bm k}\equiv \xi_{\bm k-e\bm A\tau_3}\tau_3$. 
Hence the variation of the order parameter reads
\begin{align}
\delta_A^2\Delta
&=
-iU\sum_{\bm k} {\rm Tr}\left[\tau_1\hat G_{\bm k}(\delta_A\hat\xi_{\bm k})\hat G_{\bm k}(\delta_A\hat \xi_{\bm k})\hat G_{\bm k}\right]^<
\nonumber
\\
&\quad
-\frac{i}{2}U\sum_{\bm k} {\rm Tr}\left[\tau_1\hat G_{\bm k}(\delta_A^2\hat \xi_{\bm k}-\delta_A^2\Delta\tau_1)\hat G_{\bm k}\right]^<.
\label{self consistent eq for Delta}
\end{align}
Here the lesser component of the products of the Green's functions should be understood by Langreth's rule [e.g., $(GG)^<=G^RG^<+G^<G^A$].
Equation (\ref{self consistent eq for Delta}) determines $\delta\Delta=\delta_A^2\Delta$ self-consistently. 
This is diagrammatically represented in the first line of Fig.~\ref{delta feynman diagram}.
One can show that the first term on the right hand side of Eq.~(\ref{self consistent eq for Delta}) vanishes
in the BCS theory.
By solving Eq.~(\ref{self consistent eq for Delta}) in the frequency domain, we end up with
\begin{align}
\delta\Delta
&=
\frac{\frac{i}{2}\alpha_1 U\sum_{\bm k,\omega}\xi_{\bm k} {\rm Tr}\left[\tau_1\hat G_{\bm k}(\omega+2\Omega)\tau_3\hat G_{\bm k}(\omega)\right]^<}
{1-\frac{i}{2}U\sum_{\bm k,\omega}{\rm Tr}\left[\tau_1\hat G_{\bm k}(\omega+2\Omega)\tau_1\hat G_{\bm k}(\omega)\right]^<}.
\label{pair-pair correlation}
\end{align}
Note that $\langle \tau_1 \hat G \tau_1 \hat G\rangle$ appearing in the denominator is the dynamical pair-pair correlation function.

The amplitude of the $2\Omega$ oscillation of $\delta\Delta$ diverges (i.e., APR occurs)
when the denominator of Eq.~(\ref{pair-pair correlation}) vanishes
due to fluctuations in the $\tau_1$ channel. 
Thus the resonance sensitively reflects the structure of the pair-pair correlation function.
We should emphasize that the $\tau_1$ fluctuation
appears without considering self-energy corrections beyond the mean-field BCS formalism.
Namely, the $\tau_1$ fluctuation is already present in the response to electromagnetic fields in the BCS regime
before we further include fluctuations by, e.g., the random phase approximation.
%{\it within the BCS formalism}. One does not have to go beyond the BCS theory (i.e., the strong-coupling regime)
%to include fluctuations by, for example, the random phase approximation, but the $\tau_1$ fluctuation
%is already present in the response to electromagnetic fields in the BCS regime.

As indicated in the second line of Fig.~\ref{delta feynman diagram}, $\delta\Delta$ can also be expressed in terms of the
$\tau_1$ vertex. Formally, the Higgs mode is defined as a pole of the $\tau_1$ vertex.
Therefore, the divergence of $\delta\Delta$ can indeed be rephrased as a resonance with the Higgs mode.

The explicit calculation for the correlation functions within the BCS theory enables one to 
write down $\delta\Delta$ (\ref{pair-pair correlation}) analytically as
\begin{align}
\delta\Delta
&=
\frac{1}{2}\alpha_1e^2A^2\Delta\left[\frac{1}{\lambda\, R\!\left(\Omega,T\right)}-1\right],
\label{delta Delta R}
%\\
%&\quad\times
%\begin{cases}
%\displaystyle
%\frac{1}{\lambda}\frac{\Omega}{\sqrt{\Delta^2-\Omega^2}}\frac{1}{\sin^{-1}\left(\frac{\Omega}{\Delta}\right)}-1 & \Omega<\Delta
%\vspace{.1cm}
%\\
%\displaystyle
%\frac{1}{\lambda}\frac{\Omega}{\sqrt{\Omega^2-\Delta^2}}\frac{1}{\cosh^{-1}\left(\frac{\Omega}{\Delta}\right)-i\frac{\pi}{2}}-1 & \Omega>\Delta
%\end{cases}
\end{align}
where the resonance function $R(\Omega,T)$ is given by
\begin{align}
R(\Omega,T)
&=
\mathcal P\int_\Delta^\infty d\omega
\frac{\Omega^2-\Delta^2}{(\Omega^2-\omega^2)\sqrt{\omega^2-\Delta^2}}
\tanh\bigg(\frac{\omega}{2T}\bigg)
\nonumber
\\
&\quad
-\frac{i\pi}{2}\theta(\Omega-\Delta)\frac{\sqrt{\Omega^2-\Delta^2}}{\Omega}\tanh\bigg(\frac{\Omega}{2T}\bigg)
\label{R function}
\end{align}
with $\mathcal P$ denoting the principal value of the integral.

In the limit of $2\Omega\to 2\Delta$ (i.e., $\Omega\to\Delta$), $R(\Omega,T)$ has an asymptotic form of
\begin{align}
R(\Omega,T)
&\sim
\begin{cases}
\displaystyle
\frac{\pi}{2}\frac{\sqrt{\Delta^2-\Omega^2}}{\Delta}\tanh\left(\frac{\Delta}{2T}\right) & \Omega\to \Delta-0
\vspace{.2cm}
\\
\displaystyle
-\frac{i\pi}{2}\frac{\sqrt{\Omega^2-\Delta^2}}{\Delta}\tanh\left(\frac{\Delta}{2T}\right) & \Omega\to \Delta+0.
\end{cases}
\end{align}
Hence $|\delta\Delta|$ diverges in this limit as
\begin{align}
|\delta\Delta|
\sim
\frac{\alpha_1e^2A^2\Delta^2}{\pi\lambda\tanh\left(\frac{\Delta}{2T}\right)}\frac{1}{|\Omega^2-\Delta^2|^{1/2}}.
\label{delta divergence}
\end{align}
Remarkably, the divergence persists at arbitrary temperatures $T<T_c$ with the fixed critical exponent $\frac{1}{2}$,
which indicates that the APR is robust against thermal fluctuations.

In the zero-temperature limit, $R(\Omega,T)$ is reduced to
\begin{align}
R(\Omega,T)
&\overset{T\to +0}{\sim}
\begin{cases}
\displaystyle
\frac{\sqrt{\Delta^2-\Omega^2}}{\Omega}\sin^{-1}\left(\frac{\Omega}{\Delta}\right) & \Omega<\Delta
\vspace{.2cm}
\\
\displaystyle
\frac{\sqrt{\Omega^2-\Delta^2}}{\Omega}\left[\cosh^{-1}\left(\frac{\Omega}{\Delta}\right)-i\frac{\pi}{2}\right] & \Omega>\Delta.
\end{cases}
\end{align}
Plugging this into Eq.~(\ref{delta Delta R}), one can see that 
it precisely reproduces the result (\ref{asymptotic}) derived in the previous section
[note that the seeming difference of the factor $\frac{1}{2}$ is due to the assumption
of $\bm A(t)=\bm A e^{-i\Omega t}$ in the present section and $\bm A(t)=\bm A \sin\Omega t$
in the previous section].

%\begin{align*}
%R(x,y)
%&\overset{y\to\infty}{\sim}
%\begin{cases}
%\displaystyle
%\frac{\pi}{4y}\sqrt{1-x^2} & 0\le x\le 1
%\vspace{.2cm}
%\\
%\displaystyle
%-\frac{i\pi}{4y}\sqrt{x^2-1} & x>1
%\end{cases}
%\end{align*}

\begin{figure}[t]
\includegraphics[width=7cm]{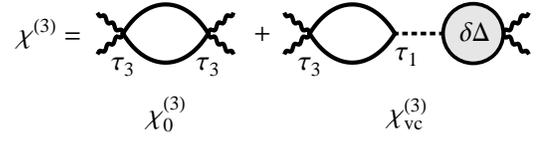}
\caption{
Feynman diagram for the THG susceptibility of superconductors.
The first and second terms on the right-hand side correspond to
$\chi_0^{(3)}$ and $\chi_{\rm vc}^{(3)}$, respectively.
}
\label{chi3 diagram}
\end{figure}

In the language of the Green's function, the current is given by
\begin{align}
j(t)=
ie\sum_{\bm k} {\rm Tr}\left[\hat v_{\bm k} \hat G_{\bm k}^<(t,t)\right],
\end{align}
where $\hat v_{\bm k}\equiv v_{\bm k-e\bm A(t)\tau_3}$. If we focus on the THG response that
is relevant to APR, we can take the third derivative with respect to $\bm A(t)$ to obtain
\begin{align}
j^{(3)}
&=
ie\sum_{\bm k} {\rm Tr}\left[(\delta_A^3 \hat v_{\bm k}) \hat G_{\bm k}^<\right]
\nonumber
\\
&\quad
+ie\sum_{\bm k} {\rm Tr}\left[(\delta_A\hat v_{\bm k})\hat G_{\bm k}(\delta_A^2\hat \xi_{\bm k}-\delta_A^2\Delta\tau_1)\hat G_{\bm k}\right].
\label{j3}
\end{align}
Here we have used the fact that odd-order derivatives of $\hat G_{\bm k}^<(t,t)$ vanish
as indicated by the pseudospin analysis [see Eqs.~(\ref{BdG1})-(\ref{BdG3})].

The first term in Eq.~(\ref{j3}) is of higher order in $(\omega_D/\epsilon_F)$ than the other terms, so that it is negligible.
Then the leading contributions to $j^{(3)}$ are the second and third terms, which we write as 
\begin{align}
j^{(3)}=\chi_{\rm tot}^{(3)}A^3=(\chi_0^{(3)}+\chi_{\rm vc}^{(3)})A^3.
\end{align}
$\chi_{\rm tot}^{(3)}$ is the total THG susceptibility, which comprises the bare susceptibility $\chi_0^{(3)}$ and vertex correction $\chi_{\rm vc}^{(3)}$.
The Feynman diagram for these is depicted in Fig.~\ref{chi3 diagram}. Here one can single out
the contribution of the $2\Omega$ collective oscillation of the order parameter to the THG signal, which is $\chi_{\rm vc}^{(3)}$.
In other words, we can distinguish the effect of quasiparticle excitations from the contribution of the Higgs mode.

We can evaluate the components of the susceptibility explicitly in the BCS theory with the function $R(\Omega,T)$ (\ref{R function}) as
\begin{align}
\chi_0^{(3)}
&=
\alpha_1^2e^4\frac{\Delta^2}{U}\left[\lambda\, R(\Omega,T)-1\right],
\\
\chi_{\rm vc}^{(3)}
&=
-\alpha_1^2e^4\frac{\Delta^2}{U}\frac{\left[\lambda\, R(\Omega,T)-1\right]^2}{\lambda\, R(\Omega,T)}.
\end{align}
Taking the sum of the two susceptibilities, we obtain the total contribution,
\begin{align}
\chi_{\rm tot}^{(3)}
&=
\chi_0^{(3)}+\chi_{\rm vc}^{(3)}
%\nonumber
%\\
%&=
=
-\alpha_1^2e^4\frac{\Delta^2}{U}
\left[\frac{1}{\lambda\, R(\Omega,T)}-1\right].
\label{chi3 total}
\end{align}
With Eqs.~(\ref{delta Delta R}) and (\ref{chi3 total}), we reproduce the previous relation (\ref{j3 Delta}).
%$j^{(3)}=\chi_{\rm tot}^{(3)}A^3=-2\alpha_1e^2 \Delta U^{-1}\delta\Delta\, A$
Note that the term $\lambda\, R(\Omega,T)$ appears in
the denominator for $\chi_{\rm tot}^{(3)}$ and in the numerator for $\chi_0^{(3)}$ in the opposite ways.

\begin{figure}[t]
\includegraphics[width=8cm]{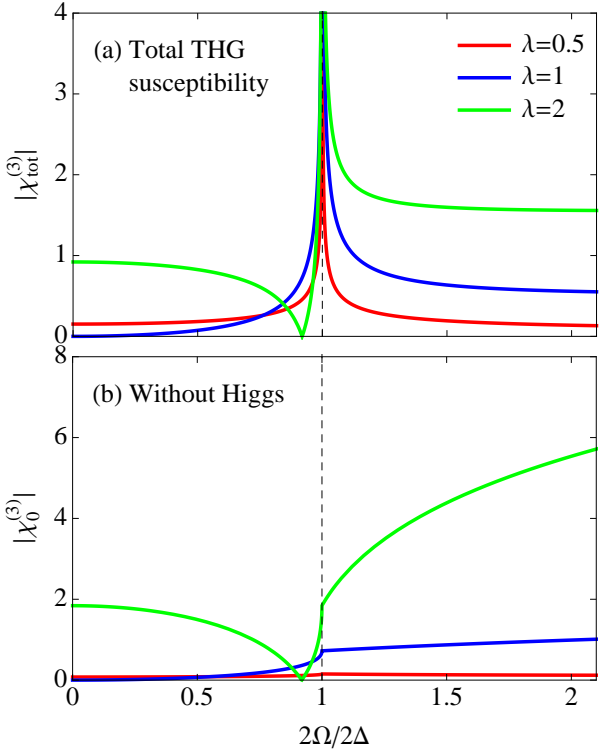}
\caption{
(Color online)
The amplitude of (a) the full and (b) bare THG susceptibilities for superconductors at $T=0$ with $\lambda=0.5, 1, 2$
in units of $\alpha_1^2e^4\omega_D^2D(\epsilon_F)$.
}
\label{chi3 fig}
\end{figure}

In Fig.~\ref{chi3 fig}, we plot $\chi_{\rm tot}^{(3)}$ along with $\chi_0^{(3)}$ for several values of $\lambda$ at $T=0$.
When we change $\lambda$, we evaluate the gap by $\Delta=\omega_D/\sinh(1/\lambda)$. We can see that $\chi_{\rm tot}^{(3)}$
diverges in a similar manner as $\delta\Delta$ (Fig.~\ref{fig:BdG}) at $2\Omega=2\Delta$, while
$\chi_0^{(3)}$ only shows kink structures. This endorses that the resonance peak is indeed a manifestation of the effect of the Higgs mode,
and cannot be explained by quasiparticle excitations or pair breaking contained in $\chi_0^{(3)}$.
As one increases $\lambda$, the amplitude of the divergence becomes larger. For $\lambda>1$, 
$\chi_{\rm tot}^{(3)}$ and $\chi_0^{(3)}$ vanish at the value of $2\Omega/2\Delta$ at which $\lambda\, R(\Omega,T)=1$ is satisfied.
Consequently, the susceptibility spectrum shows a sharp dip structure. 
This can be exploited to experimentally discern whether $\lambda>1$ or not.
We also notice that the shape of the resonance peak is significantly asymmetric about $2\Omega=2\Delta$,
which becomes more prominent for larger $\lambda$.
The asymmetry originates from the mixing of the collective mode (discrete level) with quasiparticle excitations (continuous levels).
While this is reminiscent of the Fano resonance, the form of the resonance function is not identical with the Fano form.
%Near the resonance at $T=0$, the $\lambda$ dependence is $e^{-1/\lambda}/\lambda$ for $\delta\Delta$
%and $(e^{-1/\lambda}/\lambda)^2$ for $\chi^{(3)}$. Both of $\delta\Delta$ and $\chi^{(3)}$ takes
%the maximal values at $\lambda=1$.

\begin{figure}[t]
\includegraphics[width=8cm]{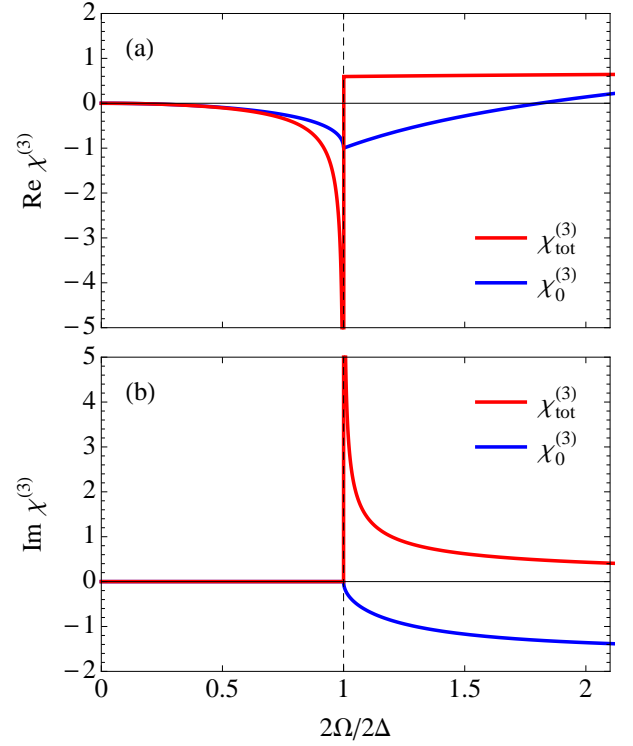}
\caption{
(Color online)
(a) The real and (b) imaginary parts of the THG susceptibility for superconductors at $T=0$ with $\lambda=1$
in units of $\alpha_1^2e^4\omega_D^2D(\epsilon_F)$.
}
\label{chi3 re im}
\end{figure}

If we look at 
the real and imaginary parts of the THG susceptibility in Fig.~\ref{chi3 re im},
the spectral features are again very different between $\chi_{\rm tot}^{(3)}$ and $\chi_0^{(3)}$.
%The most striking difference is that 
${\rm Re}\,\chi_{\rm tot}^{(3)}$ and
${\rm Im}\,\chi_{\rm tot}^{(3)}$ diverge as $\Omega\to\Delta-0$ and $\Omega\to\Delta+0$, respectively,
whereas ${\rm Re}\,\chi_0^{(3)}$ remains finite and ${\rm Im}\,\chi_0^{(3)}$ vanishes in these limits.
Both $\chi_{\rm tot}^{(3)}$ and $\chi_0^{(3)}$ have zero imaginary parts at $\Omega<\Delta$.
This simply reflects that photo-absorption is not allowed with frequencies below the energy gap
even in the nonlinear-response regime.

\begin{figure}[t]
\includegraphics[width=8cm]{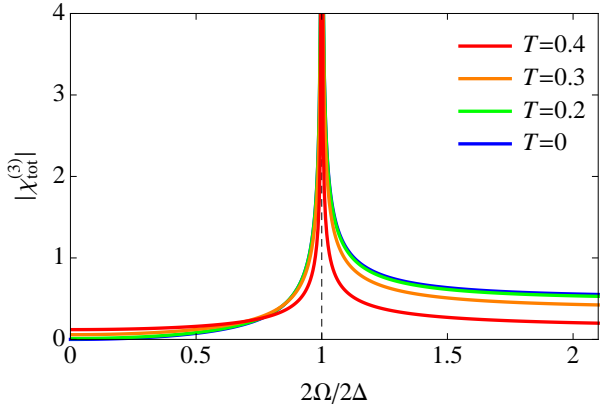}
\caption{
(Color online)
The temperature dependence of the THG susceptibility for superconductors with $\lambda=1$
in units of $\alpha_1^2e^4\omega_D^2D(\epsilon_F)$.
Here, $T_c=0.446$.
}
\label{chi3 temperature}
\end{figure}

If we turn to the temperature dependence of the THG
susceptibility in Fig.~\ref{chi3 temperature}, 
$\chi_{\rm tot}^{(3)}$ diverges for temperatures $T<T_c$, similarly to the behavior of $\delta\Delta$ (\ref{delta divergence}).
The shape of the resonance peak
does not change significantly against temperature. The insensitivity of the THG signal against temperature
should facilitate experiments where a scan of the pump frequency $\Omega$ is difficult: 
one can instead scan temperature to change $\Delta$ for a fixed $\Omega$. In Ref.~\onlinecite{Matsunaga2014},
the THG resonance peak was in fact mapped out in this way.

%\section{Effect of impurity scattering}

\section{Effect of electron-electron scattering}

So far, the argument has been based on the pairing Hamiltonian (\ref{pairing Hamiltonian}), 
which has the long-range interaction in real space.
%the treatment performed here exact.  
For more realistic models of superconductivity with short-range interactions,
the analysis above is considered to be a static mean-field approximation,
whose validity is restricted to the weak-coupling regime. 
Furthermore, the equation of motion (\ref{Bloch equation}) does not involve thermalization processes, 
which correspond to changes in the pseudospin length $|\sigma_{\bm k}|$ 
due to correlation effects.
%while the size is conserved in the treatment above with 
%Eq.~(\ref{Bloch equation}).
%If one takes account of correlation effects beyond that, the Higgs mode may acquire a finite lifetime, and the resonance width of APR will be broadened.
Thus let us go beyond the static mean field by considering the attractive Hubbard model with a driving ac field,
\begin{align}
H_{\rm Hubbard}=\sum_{\bm k\sigma}\epsilon_{\bm k-e\bm A(t)}c_{\bm k\sigma}^\dagger c_{\bm k\sigma}-U\sum_i c_{i\uparrow}^\dagger c_{i\uparrow}c_{i\downarrow}^\dagger c_{i\downarrow},
\end{align}
where $i$ labels the lattice sites and $U$ is an attractive Hubbard interaction. 
We take, as an example, a one-dimensional dispersion $\epsilon_k=-2\cos k$ with the bandwidth $W=4$ and $\alpha_1=-1$
(later in this section we also consider an infinite-dimensional lattice). 
%Other details of the lattice structure do not affect the results.
We calculate the time evolution by means of the nonequilibrium 
dynamical mean-field theory (DMFT)
\cite{FreericksTurkowskiZlatic2006,noneqDMFTreview,note}, which is extended here to the Nambu formalism for treating superconductors.
For an impurity solver for DMFT, we employ the third-order 
%weak-coupling 
perturbation theory \cite{TsujiWerner2013}, which is supposed to be reliable 
in the region $U<W$. The system is set at half filling with 
$U=3.5$, which belongs to a strong-coupling regime ($2\Delta_{T=0}/T_c\approx 5.0$ well above the BCS value). 
%Note that this model shows a coherent Higgs amplitude mode in this interaction regime \cite{TsujiEcksteinWerner2013}.

\begin{figure}
\includegraphics[width=6cm]{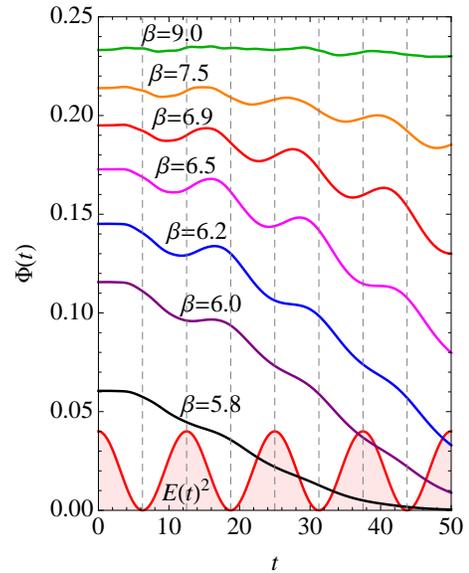}
\caption{(Color online) Temporal evolution of the superconducting order parameter $\Phi(t)$ calculated with the nonequilibrium DMFT
for the attractive Hubbard model with the 1D density of states at half filling driven by an ac field with $U=3.5$, $A=0.15$, and $\Omega=2\pi/25$ for several temperatures ($\beta^{-1}$) for the initial states.
%($\beta=5.8, 6.0, 6.2, 6.5, 6.9, 7.5, 9.0$ from bottom to top). 
The sinusoidal curve represents $\bm E(t)^2\propto \cos^2\Omega t$.
Dashed lines are a guide to the eye.}
\label{fig:dmft}
\end{figure}

\begin{figure}[t]
\includegraphics[width=8cm]{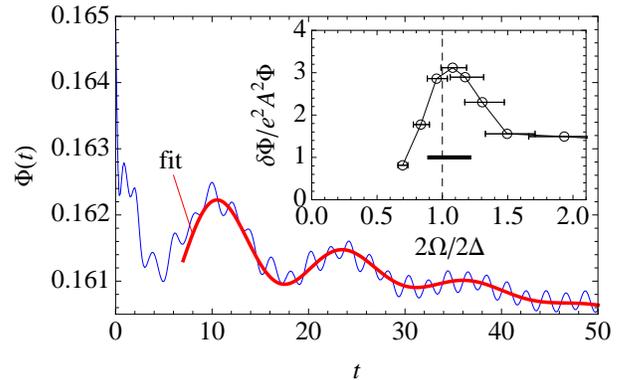}
\caption{(Color online) Temporal evolution of the order parameter $\Phi(t)$ after a quench $U\to U-\delta U$ at $t=0$ 
with $U=3.5$ and $\delta U=0.01$ in the attractive Hubbard model at $\beta=6.4$.
The rapid oscillation comes from a band-edge effect, while the slower one corresponds to the Higgs mode.
Thick (red) curve is a fit (see text).
Inset: The amplitude of the $2\Omega$ oscillating component of
the order parameter $\delta\Phi$ for the attractive Hubbard model driven by an ac field with $U=3.5$, $\Omega=2\pi/25$, and various $\Delta$. 
The bar shows the width estimated from the lifetime of the Higgs mode.}
\label{fig:dmft-resonance}
\end{figure}

The time evolution of the local superconducting order parameter, 
$\Phi(t)=\langle c_\uparrow^\dagger c_\downarrow^\dagger\rangle$, for various 
initial temperatures ($\beta^{-1}$) is shown in Fig.~\ref{fig:dmft}.
With increased total energy due to the continuous excitation, the overall value of the order parameter gradually decreases.
On top of that, the coherent oscillation of the order parameter with frequency $2\Omega$ emerges [with the same oscillation period of $\bm E(t)^2$ shown in Fig.~\ref{fig:dmft}].
The oscillation is particularly enhanced around $\beta=6.5$, and becomes invisible for $\beta=9.0$. 
The phase-shift anomaly is not clearly observed in this interaction regime.
We evaluate the energy gap $2\Delta$ in equilibrium from the single-particle spectral function $A(\omega)$,
which is calculated by Fourier transformation of the real-time simulation. 
If we measure the amplitude of the $2\Omega$ oscillation of the order parameter,
$\delta\Phi$, at the third cycle, we can clearly see in the inset of Fig.~\ref{fig:dmft-resonance} 
that a resonance peak indeed emerges at $2\Omega=2\Delta$ (the error bars represent inaccuracy in measuring $\Delta$).
The peak position corresponds to $\beta\approx 6.4$.
The result indicates that APR indeed exists beyond the static mean-field level.

However, we do notice a deviation from the BCS result; i.e., the resonance has a finite width (the inset of Fig.~\ref{fig:dmft-resonance}).
There are several factors that determine the resonance width. 
Besides extrinsic experimental factors such as the limited measurement 
time scale or energy dissipation to external environment 
(which is absent in our calculations), 
one intrinsic factor is the finite lifetime $\tau$ of the Higgs amplitude mode,
which can decay into individual excitations [note that 
Higgs does not decay into the Nambu-Goldstone (NG) mode in charged superconductors, since the energy of the NG mode is lifted to the plasma frequency, at least away from the critical regime near $T=T_c$].  
If the Higgs mode decays exponentially,
the poles $s=\pm 2i\Delta$ acquire a real part on the complex plane (Fig.~\ref{fig:contour}), 
and are thus prevented from meeting the branching points $\pm 2i\Omega$,
resulting in broadening of the resonance peak.
We can numerically evaluate the decay rate by generating the Higgs mode at $\beta=6.4$ with a small perturbation 
(here we use an interaction quench \cite{TsujiEcksteinWerner2013}),
where $\Phi(t)$ is fitted with $\Phi_0 e^{-t/\tau}\cos(2\Delta t+\theta)$ on top of a linear drift (Fig.~\ref{fig:dmft-resonance}). 
A rapid oscillation in Fig.~\ref{fig:dmft-resonance} comes from the divergence of the 1D density of states at band edges, and is irrelevant to the Higgs mode.
%simulate the Higgs mode induced by a pulse electric field, and estimate the decaying constant.
From the derived $\tau$, we estimate the resonance width as 
indicated by the bar in the inset of Fig.~\ref{fig:dmft-resonance}, which roughly coincides
with the peak width of APR with the background subtracted (Fig.~\ref{fig:dmft-resonance}).

\begin{figure}[t]
\includegraphics[width=6cm]{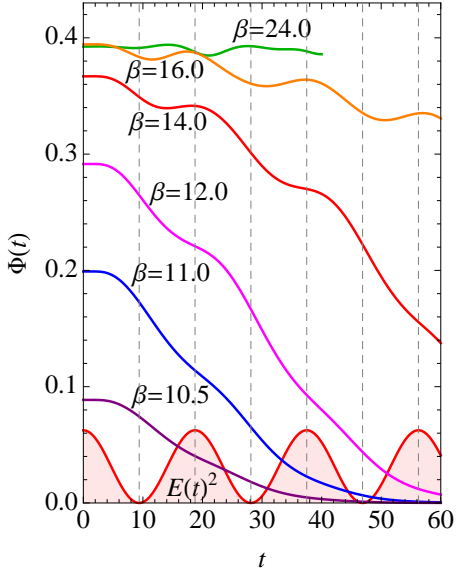}
\caption{
(Color online) Temporal evolution of the superconducting order parameter $\Phi(t)$ calculated with the nonequilibrium DMFT
for the attractive Hubbard model with the infinite-dimensional (Gaussian) density of states at half filling driven by the ac field with $U=2.25$, $A=0.2$, and $\Omega=2\pi/37.5$,
for several temperatures ($\beta^{-1}$) for the initial states.
%($\beta=5.8, 6.0, 6.2, 6.5, 6.9, 7.5, 9.0$ from bottom to top). 
The sinusoidal curve represents $\bm E(t)^2\propto \cos^2\Omega t$.
Dashed lines are a guide to the eye.
}
\label{dmft gaussian}
\end{figure}

While we have applied the nonequilibrium DMFT to a system with 1D density of states for simplicity,
we can actually confirm that the result does not change qualitatively for the infinite-dimensional hypercubic lattice with the Gaussian density of states
$D(\epsilon)=e^{-\epsilon^2}/\sqrt{\pi}$, where the DMFT formalism is no longer an approximation but becomes exact.
Let us consider the hypercubic lattice with the electric field applied along the diagonal direction,
$\bm A(t)=A(t)(1,1,1,\dots)$. The energy dispersion reads\cite{TurkowskiFreericks2005,noneqDMFTreview}
\begin{align}
\epsilon_{\bm k-\bm A(t)}
&=
\epsilon_{\bm k}\cos A(t)+\bar\epsilon_{\bm k}\sin A(t),
\end{align}
where $\epsilon_{\bm k}=-1/\sqrt{d}\sum_{i=1}^d \cos k_i$ and
$\bar\epsilon_{\bm k}=-1/\sqrt{d}\sum_{i=1}^d \sin k_i$.
%Technically, when one considers electric fields in the infinite dimensional lattice in the nonequilibrium DMFT,
This makes the momentum summation of the lattice Green's function in the nonequilibrium DMFT
a double integral with respect to $\epsilon$ and $\bar\epsilon$.
The double integral becomes computationally very heavy, especially in the present case where
we have to keep track of the system evolving over a long enough interval to capture the slow order-parameter dynamics.
To overcome the difficulty here we make use of the following formula,
\begin{align}
&\sum_{\bm k} G_{\bm k}(t,t')
\notag
\\
&=
\int d\epsilon d\bar\epsilon D(\epsilon)D(\bar\epsilon)
(i\partial_t+\mu-\epsilon\cos A-\bar\epsilon\sin A-\Sigma)^{-1}
\notag
\\
&=
\frac{1}{2}\bigg[
\int d\epsilon D(\epsilon)
\left(i\partial_t+\mu-\epsilon\cos A-\frac{1}{\sqrt{2}}\sin A-\Sigma\right)^{-1}
\notag
\\
&\quad
+\int d\epsilon D(\epsilon)
\left(i\partial_t+\mu-\epsilon\cos A+\frac{1}{\sqrt{2}}\sin A-\Sigma\right)^{-1}
\bigg]
\notag
\\
&\quad
+O(A^3),
\label{A2 formula}
\end{align}
to reduce the double integral to a single one, where
$\mu$ is the chemical potential, $\Sigma$ is the self-energy,
and we have used $\int d\bar\epsilon D(\bar\epsilon)\bar\epsilon^2=1/2$.
The formula is valid up to the second order in $A$, which is sufficient for the present purpose, since our interest is 
in the order-parameter oscillation arising from the second-order nonlinear effect.
An advantage of the above formula is that the first and second terms are in the form of the Green's function
with $\bar\epsilon$ replaced by $\pm 1/\sqrt{2}$, so that the implementation is straightforward.
We also remark that keeping the form of the Green's function is vital for maintaining the numerical stability.
As an impurity solver for the nonequilibrium DMFT for the hypercubic lattice, 
here we employ the second-order iterative perturbation theory\cite{TsujiWerner2013} to further reduce the computational cost.
If we look at the time evolution of the order parameter in the infinite-dimensional hypercubic lattice in Fig.~\ref{dmft gaussian},
the $2\Omega$ oscillation of the order parameter is prominent around $\beta=14.0$ and $16.0$,
which is close to the resonance condition $2\Omega(=4\pi/37.5=0.335)=2\Delta$($\approx 0.32$, evaluated from the 
nonequilibrium DMFT calculation for the spectral function). 
Away from this, the oscillation tends to be suppressed
and the oscillation becomes incoherent. 
This shows that APR also occurs in the nonequilibrium DMFT calculation
for the infinite-dimensional lattice where DMFT becomes exact. 
Thus the essential features of APR do not depend on a particular form of the density of states.

\section{Summary}

To summarize, we theoretically propose a phenomenon that may be called Anderson pseudospin resonance (APR) for a superconductor
driven by an ac electric field, which is confirmed 
by solving the equation of motion analytically within the BCS approximation, and by solving
the attractive Hubbard model via the nonequilibrium DMFT. 
APR can be distinguished from quasiparticle excitations or pair breaking processes
near the superconducting gap energy by looking at the divergent enhancement of
third harmonic generation.
APR provides not only a new pathway of controlling superconductors,
but also provides an avenue
offering information about dynamical aspects of the order parameter and the Higgs mode in superconductors.
Important future problems include whether APR occurs for other pairing symmetries such as the anisotropic $d$-wave pairing. 
%and how far one can pursue an analogy between the pseudospins and real spins 
%by engineering the nonlinear optical techniques.

We wish to thank R. Shimano, R. Matsunaga, H. Fujita, and A. Sugioka for stimulating discussions that motivate the present project and providing their experimental data. 
We were supported by a Grant-in-Aid for Scientific Research from MEXT 
(No.~26247057), 
and N.T. by Grant-in-Aid for Scientific Research from JSPS (No.~25104709 and No.~25800192). 
%on Innovative Areas ``Materials Design through Computics: Complex Correlation and Non-equilibrium Dynamics'' (Grant No.~25104709), and a Grant-in-Aid for Young Scientists (B) (Grant No.~25800192).

\bibliographystyle{apsrev}
\bibliography{anderson-pseudospin-resonance}

\begin{thebibliography}{34}
\bibitem[{not()}]{note}
\bibinfo{note}{Note that what essentially matters in the DMFT calculation here
  is the density of states near the Fermi energy and the $\alpha_1$ parameter.
  Thus, one may take our model adopted here as a certain infinite-dimensional
  lattice model that has the corresponding density of states and $\alpha_1$, to
  which DMFT can be applied.}
\end{thebibliography}

\end{document}